\documentclass[journal=jctcce,manuscript=article]{achemso}

\usepackage{amsmath}
\usepackage{amssymb}
\usepackage{multirow}
\usepackage[hidelinks]{hyperref}
\newcommand*{\angstrom}{Å}
\newcommand*{\degree}{^\circ}

\usepackage{tikz}
\tikzstyle{wire}=[thick]

\title{Moments-based quantum computation of the electric dipole moment of molecular systems}
\author{Michael A. Jones}
\affiliation[University of Melbourne]{School of Physics, University of Melbourne, Parkville 3010, Australia}
\alsoaffiliation[QUBIC]{ARC Centre of Excellence in Quantum Biotechnology, University of Melbourne, Parkville, 3010, Australia}
\email{jonesm3@student.unimelb.edu.au}
\author{Harish J. Vallury}
\affiliation[University of Melbourne]{School of Physics, University of Melbourne, Parkville 3010, Australia}
\alsoaffiliation[CSIRO]{CSIRO Data61, Marsfield 2122, Australia}
\author{Manolo C. Per}
\affiliation[CSIRO]{CSIRO Data61, Clayton 3168, Australia}
\alsoaffiliation[QUBIC]{ARC Centre of Excellence in Quantum Biotechnology, University of Melbourne, Parkville, 3010, Australia}
\author{Harry M. Quiney}
\affiliation[University of Melbourne]{School of Physics, University of Melbourne, Parkville 3010, Australia}
\author{Lloyd C. L. Hollenberg}
\affiliation[University of Melbourne]{School of Physics, University of Melbourne, Parkville 3010, Australia}
\alsoaffiliation[QUBIC]{ARC Centre of Excellence in Quantum Biotechnology, University of Melbourne, Parkville, 3010, Australia}

\keywords{quantum computing,quantum algorithms,electronic structure,electric dipole}

\begin{document}


\begin{abstract}
With rapid progress being made in the development of platforms for quantum computation, there has been considerable interest in whether present-day and near-term devices can be used to solve problems of relevance. A commonly cited application area is the domain of quantum chemistry. While most experimental demonstrations of quantum chemical calculations on quantum devices have focused on the ground-state electronic energy of the system, other properties of the ground-state, such as the electric dipole moment, are also of interest. Here we employ the quantum computed moments (QCM) method, based on the Lanczos cluster expansion, to estimate the dipole moment of the water molecule on an IBM Quantum superconducting quantum device. The noise-mitigated results agree with full configuration interaction (FCI) calculations to within 0.03 $\pm$ 0.007 debye (2\% $\pm$ 0.5\%), compared to direct expectation value determination (i.e. VQE) with errors on the order of 0.07 debye (5\%), even when the VQE calculation is performed without noise.
This demonstrates that moments-based energy estimation techniques can be adapted to noise-robust evaluation of non-energetic ground-state properties of chemical systems.
\end{abstract}

\section{Introduction}
The simulation of quantum chemical systems seems a natural application of interest for quantum computers. However, due to strict accuracy requirements and the need for deep circuits, practical demonstrations of quantum-computed chemistry algorithms have been limited to proof-of-principle experiments, usually employing the variational quantum eigensolver (VQE) \cite{PeruzzoMcClean_VQE_2014} to estimate the ground-state energy of small systems. More recently, progress has been made on the implementation of more sophisticated ground-state energy estimation methods; such as those based on 
variations of the quantum phase estimation algorithm \cite{Li_ResonantTransitions_2019},
imaginary time evolution \cite{Yeter-Aydeniz_QITEQLanczos_2020,ZongHuaiCai_QITE_2023},
Hamiltonian moments \cite{Suchsland_VariationalMoments_2021,Jones_Hydrogen_2022,GuoSunQianGong_USTC_2023,Jones_Water_2024,Per_Helium_2025},
subspace expansion \cite{Colless_QSE_2018,Motta_QSE_2023,Castellanos_EFQSE_2023},
and selected configuration interaction (also known as sample-based quantum diagonalisation) \cite{Kanno_QSCI_2023,RobledoMoreno_N2QSCI_2024,Kaliakin_SQD_2024,Liepuoniute_SQD_2024,PellowJarman_HIVQE_2025,Yu_QSCI_2025}. Additionally, other properties of interest have been computed, such as excited states and band-structure calculations \cite{Colless_QSE_2018,Gao_TADF_2021,Gocho_VQEAC_2023,Castellanos_EFQSE_2023,Ollitrault_EquationOfMotion_2020,Ganzhorn_VQE_2019,Cerasoli_VQD_2020,Ohgoe_QSE_2024,Singh_H2NMR_2024,Sureshbabu_MLBandStructure_2021,Zhang_BandStructure_2024,Liepuoniute_SQD_2024},
vibrational modes \cite{Stober_VibrationalModes_2022}
and dipole moments \cite{Rice_Batteries_2021,Chawla_RelDipoles_2024,DeLima_1RDM_2025}.
Dipole moments and similar properties that do not depend on excited electronic states or motion of the atomic nuclei seem a straightforward extension to the ground-state energy estimation problem. To date, however, quantum hardware computation of these properties has been primarily restricted to direct expectation value determination, rather than taking advantage of techniques to improve energy estimation such as those listed above. However, we note that the classical description obtained through sample-based methods makes this extension straightforward and recent work by Oumarou et. al. \cite{Oumarou_KrylovProperties_2025} has proposed methods to estimate such properties through Krylov subspace diagonalisation.
Additionally, while specific methods have been proposed for the estimation molecular response properties \cite{Cai_ResponseProperties_2020,Huang_ResponseProperties_2022,Kumar_ResponseProperties_2023,Sun_iToffoli_2024} i.e. the response of properties such as the dipole moment to external fields due to excitation to higher energy states, including quantum hardware computations \cite{Huang_ResponseProperties_2022,Sun_iToffoli_2024}, our interest is in evaluating non-energetic properties of the ground-state in the absence of external potentials and in improving the accuracy of results when access to the exact ground-state is infeasible.

Here, we show improvement over the direct estimation of ground-state observables, using the electric dipole moment of the water molecule as an example, through use of the Hamiltonian moments and the quantum computed moments (QCM) method \cite{Vallury_QCM_2020,Vallury_NoiseRobust_2023}, which does not require any additional quantum circuit depth. Based on the Lanczos cluster expansion framework \cite{Hollenberg_PlaquetteExpansion_1993,HollenbergWitte_Nonperturbative_1994,HollenbergWitte_AnalyticSolution_1996}, the noise-robust, moments-corrected estimate utilises a Hellmann-Feynman approach \cite{Hollenberg_Magnetisation_1994,Witte_Magnetisation_1997,Kassal_MolecularProperties_2009,Vallury_Observables_2023}, and was previously applied to Heisenberg models in simulation \cite{Vallury_Observables_2023}. Here, we extend this by demonstrating implementation of the method using results from an IBM superconducting quantum device \cite{Qiskit_2024}. This provides experimental verification that the moments-based QCM method produces superior estimates for such quantities compared to direct evaluation when applied to chemical systems. In addition we consider a number of ways in which the Hellmann-Feynman approach can be combined with error-mitigation techniques and investigate their behaviour in Appendix \ref{app:noise}.

Moments-based approaches are closely related to quantum Krylov methods \cite{Parrish_QFD_2019,Stair_MRSQK_2020,Yoshioka_KQD_2025,Yu_QSCI_2025} with moments-based techniques corresponding to a specific choice of the Krylov subspace. In particular, the Lanczos cluster expansion \cite{Hollenberg_PlaquetteExpansion_1993,HollenbergWitte_Nonperturbative_1994,HollenbergWitte_AnalyticSolution_1996} based correction used here essentially results in an analytic diagonalisation of the subspace including approximate contributions from the remainder of the Hilbert space.
However, our technique and that proposed by Oumarou et. al.\cite{Oumarou_KrylovProperties_2025} are quite different. Where we accept a higher measurement cost to minimise the quantum circuit depth, Oumarou et. al. \cite{Oumarou_KrylovProperties_2025} utilise quantum signal processing techniques to reduce the measurement cost at the expense of deeper circuits.

\section{Method}

\subsection{Quantum computed moments for ground-state energy estimation}
\begin{figure}
    \includegraphics{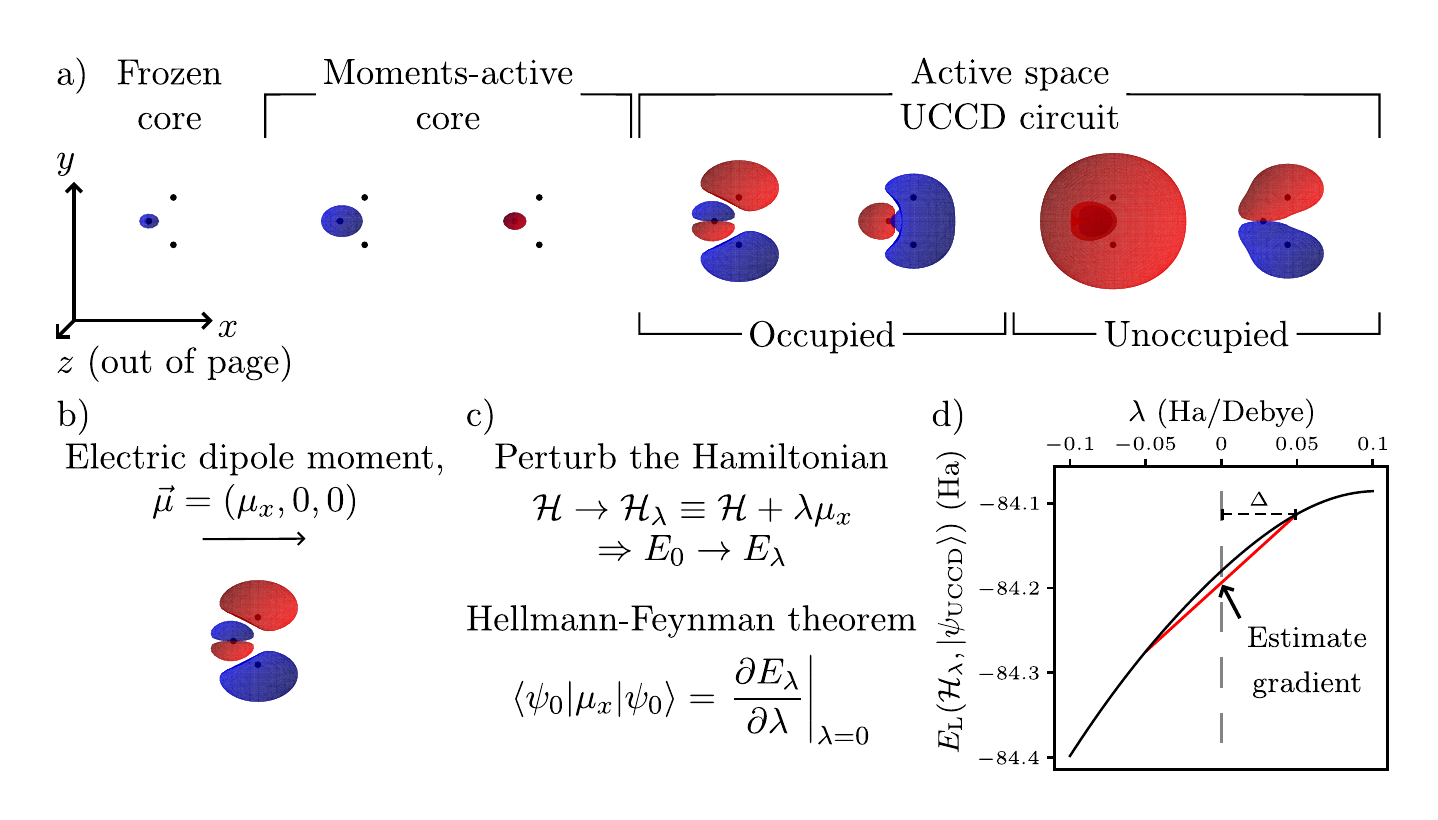}
    \caption{
    Details of the application of the method to the electric dipole moment of the water molecule. a) The STO-3G molecular orbitals used to represent the water molecule. b) Due to symmetry considerations, the electric dipole moment must be aligned with the $x$-axis. We use the convention that the dipole moment vector points from negative to positive charge. c) By perturbing the Hamiltonian with the dipole operator, the Hellmann-Feynman theorem can be applied to compute the dipole moment. d) $E_\lambda$ is approximated by $E_\mathrm{L}(\mathcal{H}_\lambda,|\psi_\mathrm{UCCD}\rangle)$ and the finite difference method is employed to estimate the derivative. Note that the trial-state $|\psi_\mathrm{trial}\rangle=|\psi_\mathrm{UCCD}\rangle$ is independent of $\lambda$, allowing for reuse of the same quantum measurements and thereby reducing the effects of stochastic error. $\Delta$ is the step-size used for the finite difference method.}
    \label{fig:1}
\end{figure}
The trial-state preparation and measurement follow the methods outlined in previous work \cite{Jones_Water_2024} and are summarised here for completeness.
The water molecule (H$_2$O) at equilibrium geometry (HOH bond angle; 104.5$\degree$, symmetric OH bond length; 0.96\angstrom) is represented in the STO-3G basis by the second-quantised Hamiltonian:
\begin{align}
    \mathcal{H}&=\sum_{jk}^{N_s}h_{jk}a^\dagger_ja_k+\sum_{jklm}^{N_s}g_{jklm}a^\dagger_ja^\dagger_ka_la_m,\label{eqn:Hamiltonian}
\end{align}
where $N_s=14$ is the number of spin-orbitals, $a^\dagger_j$ (and $a_j$) are the electronic creation (and annihilation) operators and $h_{jk}$ (and $g_{jklm}$) are the classically computed one- (and two-) body integrals. The spatial orbitals are shown in Figure \ref{fig:1}a, each corresponds to a pair of spin-orbitals that account for the electronic spin degree of freedom. Subsequently, the core (1s) orbitals of the oxygen atom are frozen, as they do not participate in the bonding. The resulting Hamiltonian also follows the form of equation \ref{eqn:Hamiltonian} with $N_s=12$ spin-orbitals, but with modified values for $h_{jk}$ and $g_{jklm}$, and an additional constant term. The Hamiltonian moment operators, $\mathcal{H}^p$, are determined using Wick's theorem to simplify products of creation and annihilation operators, for $p\in\{1,2,3,4\}$. Each of these moment operators is then reduced to an 8 spin-orbital operator by freezing 2 additional orbitals (predominantly the oxygen 2s and 2p$_z$ orbitals, where $z$ is the axis perpendicular to the plane of the molecule). As a result, these orbitals are not directly simulated by the quantum device but are incorporated by the moments-based correction. The remaining 8 spin-orbitals are each mapped to a qubit via the Jordan-Wigner encoding \cite{JordanWigner_JWE_1928} and a selected Unitary-Coupled-Cluster-Doubles (UCCD, \cite{Taube_UCC_2006}) trial-state, $|\psi_\mathrm{UCCD}\rangle$, is constructed as a variational wavefunction which is measured in 200 different bases (25,000 measurements each) using readout-error mitigation (via the M3 package\cite{Nation_M3_2021}) and symmetry verification \cite{BonetMonroig_SV_2018} (based on the total spin and spin-projection) to construct the 4-body Reduced Density Matrix (RDM). Since there are only 4 electrons remaining in the active-space, the 4-RDM is sufficient to evaluate all relevant operators. The RDM is then rescaled \cite{Tilly_QRDM_2021} to enforce the correct trace.
The circuit used to prepare the trial-state has 8 qubits, 22 two-qubit gates and a depth of 25 gates. Each measurement basis requires a different configuration of additional one- and two-qubit gates.

In our previous work\cite{Jones_Water_2024}, the Hamiltonian moments for the water molecule are evaluated from the 4-RDM for both the trial-state and the Hartree-Fock reference state, $|\psi_\mathrm{HF}\rangle$, with the latter used to mitigate noise in the former by assuming a depolarising noise model \cite{Dalzell_WhiteNoise_2021},
\begin{align}
    \mathrm{Tr}(\rho^\mathrm{noisy}\mathcal{H}^p)&=(1-q_p)\mathrm{Tr}(\rho^\mathrm{exact}\mathcal{H}^p)-q_p\mathrm{Tr}(\rho^\mathrm{mixed}\mathcal{H}^p).\label{eqn:NoiseModel}
\end{align}
Using the reference state, which has an efficiently computed classical description, $\rho^\mathrm{exact}_\mathrm{HF}$, the effective noise level, $q_p$, of the circuit can be found for each moment and Equation \ref{eqn:NoiseModel} can be inverted to correct the noisy trial-state result \cite{Czarnik_CDR_2021,LolurSkogh_ReferenceStates_2023}.

The moments are then combined in an analytic formula  derived from the cluster expansion of the Lanczos recursion \cite{Hollenberg_PlaquetteExpansion_1993}, that sums all Lanczos coefficients (truncated to fourth order in the moments) to obtain a corrected ground-state energy estimate, $E_\mathrm{L}$\cite{HollenbergWitte_Nonperturbative_1994,HollenbergWitte_AnalyticSolution_1996};
\begin{align}
    E_\mathrm{L}&\equiv c_1-\frac{c_2^2}{c_3^2-c_2c_4}\left(\sqrt{3c_3^2-2c_2c_4}-c_3\right),\label{eqn:EL}\\
    c_p&=\langle\mathcal{H}^p\rangle-\sum_{j=0}^{p-2}\binom{p-1}{j}c_{j+1}\langle \mathcal{H}^{p-1-j}\rangle.\nonumber
\end{align}
We note that quantum computation of Hamiltonian moments has also been considered in other contexts \cite{Seki_QPM_2021,Suchsland_VariationalMoments_2021}, most notably in relation to the connected moments expansion \cite{Cioslowski_CMX_1987,Kowalski_CMX_2020,Peng_VariationalPDS_2021,Claudino_CMX_2021,Ganoe_CMX_2023,GuoSunQianGong_USTC_2023}.
However, the combination of error-mitigation and moments-based correction described here was seen to work well, giving ground-state energy estimates within millihartree accuracy of the FCI result \cite{Jones_Water_2024}. Here, we aim to reproduce this accuracy for non-energy observables (namely the electric dipole moment).

\subsection{Quantum computed moments for arbitrary ground-state observable estimation}\label{sec:observables}
To estimate ground-state observables via the Hellmann-Feynman theorem in the context of moments-based methods \cite{Hollenberg_Magnetisation_1994,Witte_Magnetisation_1997,Vallury_Observables_2023}, consider the modified Hamiltonian,
\begin{align}
    \mathcal{H}_\lambda&\equiv\mathcal{H}+\lambda\mu,\label{eqn:ModifiedHamiltonian}
\end{align}
where $\lambda$ is a real scalar and $\mu$ is the observable of interest. For concreteness, let $\mu=\mu_x$ be the $x$-component of the electric dipole moment operator (the only component with non-vanishing expectation value by symmetry, Figure \ref{fig:1}b), 
\begin{align}
    \mu&=\sum_{jk}^{N_s}f_{jk}a^\dagger_ja_k,\label{eqn:Dipole}
\end{align}
where $f_{jk}$ are classically computed integrals.
Let $|\Phi_\lambda\rangle$ be the ground-state of $\mathcal{H}_\lambda$ with energy $E_\lambda$, then the Hellmann-Feynman theorem states:
\begin{align}
    \frac{\partial E_\lambda}{\partial\lambda}
    &=
    \langle\Phi_\lambda|\mu|\Phi_\lambda\rangle,
\end{align}
and evaluating at $\lambda=0$ gives expectation values with respect to the ground-state of the unmodified Hamiltonian
\begin{align}
    \left.\frac{\partial E_\lambda}{\partial\lambda}\right|_{\lambda=0}&=\langle\Phi_0|\mu|\Phi_0\rangle
    \equiv\mu_0
    \label{Eqn:HellmannFeynman}.
\end{align}
While the Hellmann-Feynman theorem also holds for states other than energy eigenstates, subject to conditions on the energy derivative with respect to trial-state parameters, here the formulation in terms of the ground-state is sufficient, as we aim to estimate non-energetic ground-state properties (right-hand side of Equation \ref{Eqn:HellmannFeynman}) from approximations to the ground-state energy (left-hand side of Equation \ref{Eqn:HellmannFeynman}). The applicability of various forms of the Hellmann-Feynman theorem is a much studied topic \cite{Epstein_HellmannFeynman_1967,McWeeny_Methods_1969}.

To approximate the derivative, we can employ the finite difference method (Figure \ref{fig:1}d), resulting in the approximation:
\begin{align}
    \mu_0
    &\approx\frac{1}{2\Delta}\left(E_\Delta-E_{-\Delta}\right).
\end{align}
Further, using the moments-based correction to improve the evaluation of $E_{\pm\Delta}$ and therefore the estimation of the dipole moment relative to the standard variational estimate gives
\begin{align}
    \mu_0&\approx\frac{1}{2\Delta}\big[E_\mathrm{L}(\lambda=\Delta)-E_\mathrm{L}(\lambda=-\Delta)\big].
\end{align}
Note that the notation used here is that $\lambda$ is a variable (with respect to which we would like to take the derivative) while $\pm\Delta$ is the value assigned to $\lambda$ to compute the finite difference approximation to the derivative.

When performing finite-difference computations on a quantum device, often the accuracy of the difference formula is severely limited by stochastic (sampling) noise. However, in this case the parameter with respect to which the derivative is taken, $\lambda$, appears in post-processing rather than in the quantum computation, which allows reuse of the quantum output for both $\lambda=\pm\Delta$ values and vastly reduces the accumulation of stochastic noise.
In the absence of the moments-based correction, reusing the trial-state this way would essentially make the approximation that the ground-state is independent of the change in the Hamiltonian, leading to systematic error due to neglecting the wavefunction response terms.
If the response terms are ignored completely and expectation values are taken at $\lambda=\pm\Delta$, then the estimate of the dipole moment is simply the expectation value of the dipole operator with respect to the trial-state,
\begin{align}
    \langle\psi_\mathrm{trial}|\mu|\psi_\mathrm{trial}\rangle
    &=\frac{\langle\psi_\mathrm{trial}|\mathcal{H}_\Delta|\psi_\mathrm{trial}\rangle-\langle\psi_\mathrm{trial}|\mathcal{H}_{-\Delta}|\psi_\mathrm{trial}\rangle}{2\Delta}.\label{Eqn:TrialExpectation}
\end{align}
While Equation \ref{Eqn:TrialExpectation} holds exactly (since the modified Hamiltonian is linear in $\lambda$ there is no error due to the finite difference method), it is not practically useful as the cost of evaluating the left- and right-hand sides are roughly equal. Additionally, the dipole-moment estimate obtained is the expectation value with respect to the trial-state, not the ground-state, so systematically approaching the ground-state expectation value requires improving the trial-state leading to additional quantum gates and therefore increased noise.

The Hellmann-Feynman theorem may be of practical use in cases where we have access to approximate ground-state energies, but not to the corresponding state.
This is the case for a number of ground-state energy estimation techniques such as the moments-based correction \cite{Hollenberg_Magnetisation_1994}. In this case, the estimates are able to incorporate additional information from the Hamiltonian to circumvent the limit imposed by the trial-state, effectively incorporating some of the wavefunction response and improving on the dipole moment estimate. We denote the dipole moment estimate computed this way using the moments-based correction, $E_\mathrm{L}$, as $\mu_\mathrm{L}(\Delta,|\psi\rangle)$, so the final expression used for the dipole-moment estimation in this work is
\begin{align}
    \mu_0&\approx\mu_\mathrm{L}(\Delta,|\psi_\mathrm{trial}\rangle)
    \equiv\frac{1}{2\Delta}\big[E_\mathrm{L}(\Delta,|\psi_\mathrm{trial}\rangle)-E_\mathrm{L}(-\Delta,|\psi_\mathrm{trial}\rangle)].
\end{align}

\begin{figure}
    \centering
    Estimation of modified moments, using $\langle\mathcal{H}_\Delta^2\rangle$ as an example
    \scalebox{1}{
    \begin{tikzpicture}[node distance=0cm]
        \node(B1)[fill=red!50,draw=black,rounded corners,minimum width=4.7cm,minimum height=1.5cm,text width=4.5cm,text centered]{\footnotesize Compute: $\mathcal{H}^2$, $\mathcal{H}\mu$, $\mu^2$};
        \node(B2)[fill=orange!50,below of=B1,yshift=-1.75cm,draw=black,rounded corners,minimum width=4.7cm,minimum height=1.5cm,text width=4.5cm,text centered]{\footnotesize Evaluate:\\$\langle\mathcal{H}^2\rangle$, $\langle\{\mathcal{H},\mu\}\rangle$, $\langle\mu^2\rangle$};
        \node(B3)[fill=blue!30,below of=B2,yshift=-1.75cm,draw=black,rounded corners,minimum width=4.7cm,minimum height=1.5cm,text width=4.5cm,text centered]{\footnotesize Assign: $\lambda=\Delta$};
        \node(B4)[fill=white!40,below of=B3,yshift=-1.75cm,draw=black,rounded corners,minimum width=4.7cm,minimum height=1.5cm,text width=4.5cm,text centered]{\footnotesize Compute: $\langle\mathcal{H}_\Delta^2\rangle=\langle\mathcal{H}^2\rangle+\Delta\langle\{\mathcal{H},\mu\}\rangle+\Delta^2\langle\mu^2\rangle$};
        \node(B5)[fill=yellow!50,below of=B4,yshift=-1.75cm,draw=black,rounded corners,minimum width=4.7cm,minimum height=1.5cm,text width=4.5cm,text centered]{\footnotesize Error mitigate: $\langle\mathcal{H}_\Delta^2\rangle$};
        \draw[wire,-stealth](B1)--(B2);
        \draw[wire,-stealth](B2)--(B3);
        \draw[wire,-stealth](B3)--(B4);
        \draw[wire,-stealth](B4)--(B5);
        
        \node(A1)[fill=blue!30,below of=B1,xshift=-5.1cm,draw=black,rounded corners,minimum width=4.7cm,minimum height=1.5cm,text width=4.5cm,text centered]{\footnotesize Assign: $\lambda=\Delta$};
        \node(A2)[fill=white!40,below of=A1,yshift=-1.75cm,draw=black,rounded corners,minimum width=4.7cm,minimum height=1.5cm,text width=4.5cm,text centered]{\footnotesize Simplify: $\mathcal{H}_\Delta=\mathcal{H}+\Delta\mu$};
        \node(A3)[fill=red!50,below of=A2,yshift=-1.75cm,draw=black,rounded corners,minimum width=4.7cm,minimum height=1.5cm,text width=4.5cm,text centered]{\footnotesize Compute: $\mathcal{H}_\Delta^2$};
        \node(A4)[fill=orange!50,below of=A3,yshift=-1.75cm,draw=black,rounded corners,minimum width=4.7cm,minimum height=1.5cm,text width=4.5cm,text centered]{\footnotesize Evaluate: $\langle\mathcal{H}_\Delta^2\rangle$};
        \node(A5)[fill=yellow!50,below of=A4,yshift=-1.75cm,draw=black,rounded corners,minimum width=4.7cm,minimum height=1.5cm,text width=4.5cm,text centered]{\footnotesize Error mitigate: $\langle\mathcal{H}_\Delta^2\rangle$};
        \draw[wire,-stealth](A1)--(A2);
        \draw[wire,-stealth](A2)--(A3);
        \draw[wire,-stealth](A3)--(A4);
        \draw[wire,-stealth](A4)--(A5);

        \node(C1)[fill=red!50,below of=B1,xshift=5.1cm,draw=black,rounded corners,minimum width=4.7cm,minimum height=1.5cm,text width=4.5cm,text centered]{\footnotesize Compute: $\mathcal{H}^2$, $\mathcal{H}\mu$, $\mu^2$};
        \node(C2)[fill=orange!50,below of=C1,yshift=-1.75cm,draw=black,rounded corners,minimum width=4.7cm,minimum height=1.5cm,text width=4.5cm,text centered]{\footnotesize Evaluate:\\$\langle\mathcal{H}^2\rangle$, $\langle\{\mathcal{H},\mu\}\rangle$, $\langle\mu^2\rangle$};
        \node(C3)[fill=yellow!50,below of=C2,yshift=-1.75cm,draw=black,rounded corners,minimum width=4.7cm,minimum height=1.5cm,text width=4.5cm,text centered]{\footnotesize Error mitigate: $\langle\mathcal{H}^2\rangle$, $\langle\{\mathcal{H},\mu\}\rangle$, $\langle\mu^2\rangle$};
        \node(C4)[fill=blue!30,below of=C3,yshift=-1.75cm,draw=black,rounded corners,minimum width=4.7cm,minimum height=1.5cm,text width=4.5cm,text centered]{\footnotesize Assign: $\lambda=\Delta$};
        \node(C5)[fill=white!40,below of=C4,yshift=-1.75cm,draw=black,rounded corners,minimum width=4.7cm,minimum height=1.5cm,text width=4.5cm,text centered]{\footnotesize Compute: $\langle\mathcal{H}_\Delta^2\rangle=\langle\mathcal{H}^2\rangle+\Delta\langle\{\mathcal{H},\mu\}\rangle+\Delta^2\langle\mu^2\rangle$};
        \draw[wire,-stealth](C1)--(C2);
        \draw[wire,-stealth](C2)--(C3);
        \draw[wire,-stealth](C3)--(C4);
        \draw[wire,-stealth](C4)--(C5);

        \node [below of=A1,yshift=1.3cm,text width=5cm,text centered]{Method A\\(\emph{de novo} moments):};
        \node [below of=B1,yshift=1.3cm,text width=5cm,text centered]{Method B\\(post-mitigated moments):};
        \node [below of=C1,yshift=1.3cm,text width=5cm,text centered]{Method C\\(pre-mitigated moments):};
        \node [below of=A5,yshift=-1.75cm,text width=4.5cm,text centered]{\small Expensive when \mbox{computing} many values of $\lambda$};
        \node [below of=B5,yshift=-1.75cm,text width=4.5cm,text centered]{\small Method of choice};
        \node [below of=C5,yshift=-1.75cm,text width=4.5cm,text centered]{\small Unpredictable in the presence of noise};
    \end{tikzpicture}}
    \caption{Three methods for evaluating the second moment, $\langle\mathcal{H}_\Delta^2\rangle$. The computational bottleneck is the implementation of Wick's theorem to multiply the fermionic operators (red boxes). Performing these steps in a way that is independent of $\Delta$ (methods B and C) allows for much more efficient evaluation of $\langle\mathcal{H}_\Delta^2\rangle$ at many values of $\lambda$, since steps prior to assigning a value of $\lambda$ (above the blue boxes) do not need to be repeated. However, for certain values of $\lambda$, method C was seen to be unstable to errors from the quantum computation (see Appendix \ref{app:noise}). Note that $\langle\{\mathcal{H},\mu\}\rangle\equiv\langle\mathcal{H}\mu+\mu\mathcal{H}\rangle=2\mathrm{Re}\langle\mathcal{H}\mu\rangle$. The methods may be generalised to evaluate $\langle\mathcal{H}_\Delta^p\rangle.$}
    \label{fig:2}
\end{figure}
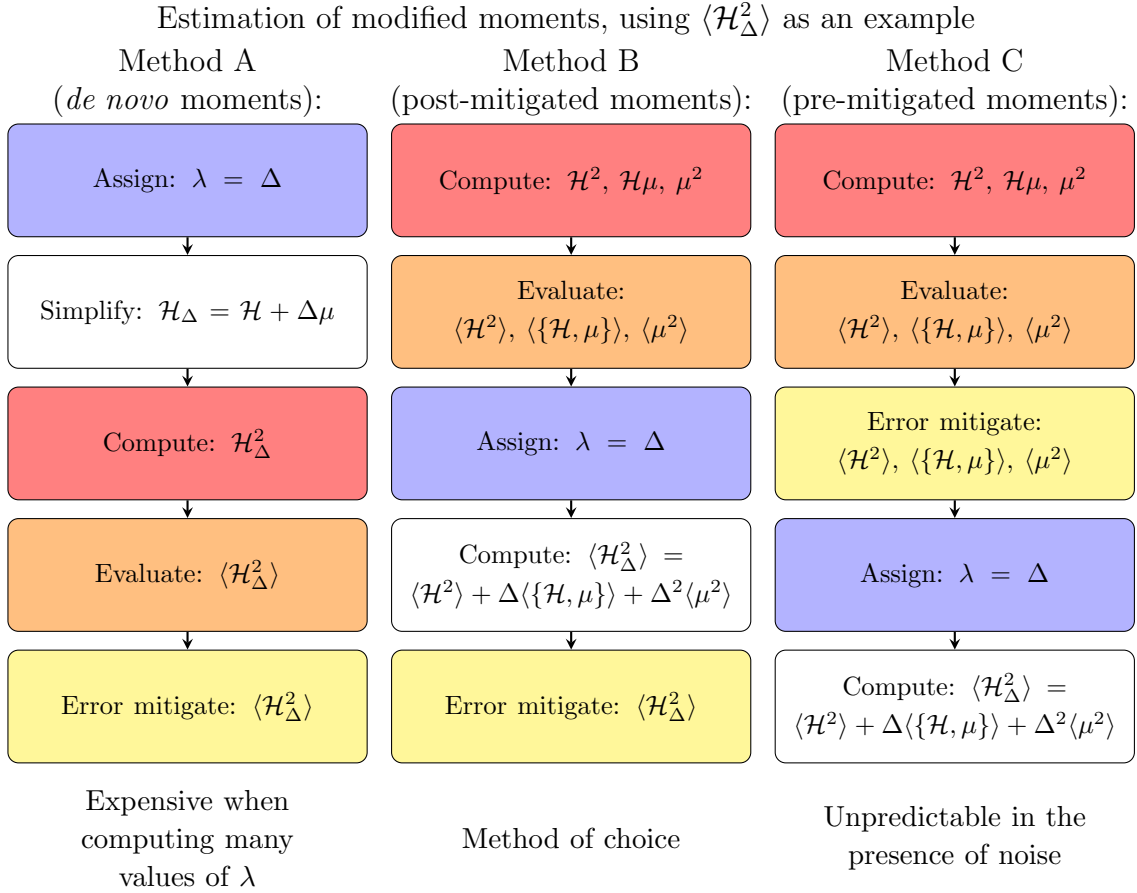

Figure \ref{fig:2} outlines three methods for calculating the reference-state calibrated moments from the reduced density matrix. The first of these (Method A), we term \emph{de novo} moments, as each time $\Delta$ is adjusted (e.g. to test for convergence to the $\Delta\rightarrow0$ limit), the classical computation of the moments must begin anew, though the quantum computation need not be repeated. To minimise the required classical computation, we can instead express $\mathcal{H}_\lambda^p$ analytically as a polynomial in $\lambda$ and compute the operator-valued coefficients (Methods B and C in Figure \ref{fig:2}). This allows evaluation of the moments-corrected ground-state energy estimate for many values of $\Delta$ with minimal additional processing. 
However, as indicated in Figure \ref{fig:2}, we apply the reference-state calibration to the modified moments, $\langle\mathcal{H}_\lambda^p\rangle$. We term this method `post-mitigated moments' as the reference-state error mitigation is performed after construction of the modified moments (Method B). In contrast to applying the reference-state calibration to the coefficients, $\langle\mathcal{H}\rangle$, $\langle\mu\rangle$ etc. (Method C), we find this to be more stable to noise. We term Method C, the `pre-mitigated moments' method, since reference-state error mitigation is performed before construction of the modified moments. Further investigation of the adverse error propagation is given in Appendix \ref{app:noise}. Based on the noise analysis, Appendix \ref{app:noise} also introduces Methods D and E which we refer to as `truncated post-mitigated moments' and `truncated pre-mitigated moments' respectively and correspond to discarding higher-order dependencies on $\Delta$.

We note that there are many improvements and adaptations that can be made to VQE to improve its accuracy and/or reduce the quantum computational cost and we expect many of these to be compatible with the moments-based methods for energy and non-energetic property estimation. Such techniques could include explicit correlation methods \cite{McArdle_Transcorrelated_2020,Motta_Transcorrelated_2020,Kumar_Transcorrelated_2022,Schleich_2R12_2022,Sokolov_Transcorrelated_2023,Dobrautz_Transcorrelated_2024}, Hamiltonian factorisation and downfolding schemes \cite{Poulin_DoubleFactorisation_2015,Lee_Hypercontraction_2021,Motta_LowRank_2021,Oumarou_DoubleFactorisation_2024,
Metcalf_Downfolding_2020,Bauman_Downfolding_2019,Bauman_Downfolding2_2019,Bauman_Downfolding3_2021,Bauman_Downfolding4_2022,Huang_Downfolding_2023}, specific basis choices or the use of basis-set-free methods \cite{Babbush_PlaneWaves_2018,Hong_Wavelets_2022,Kottmann_BasisSetFree_2021} and/or perturbative techniques \cite{Ryabinkin_Perturbative_2021,Tammaro_NEVPT_2023}.

We also note that the major computational bottlenecks of the method presented here are the required number of quantum measurements and the necessary classical pre-computation. The latter can be circumvented by methods which encode the Hamiltonian directly into the quantum circuit \cite{Duan_MPS_2015,Seki_QPM_2021}. Such methods may also be able to reduce the shot requirement. However, using such methods, the Hamiltonian cannot be varied in post-processing as it is encoded directly into the quantum circuit, this would likely result in a loss of robustness in the finite differencing step used to estimate non-energetic properties, as the stochastic (sampling) errors on each side of the finite difference would be independent. Whether a ``best of both'' solution, that reduces the pre-processing and the number of shots while avoiding amplification of stochastic noise in the finite difference step, remains an open question.

\section{Results and Discussion}
\begin{figure}
    \centering
    \includegraphics{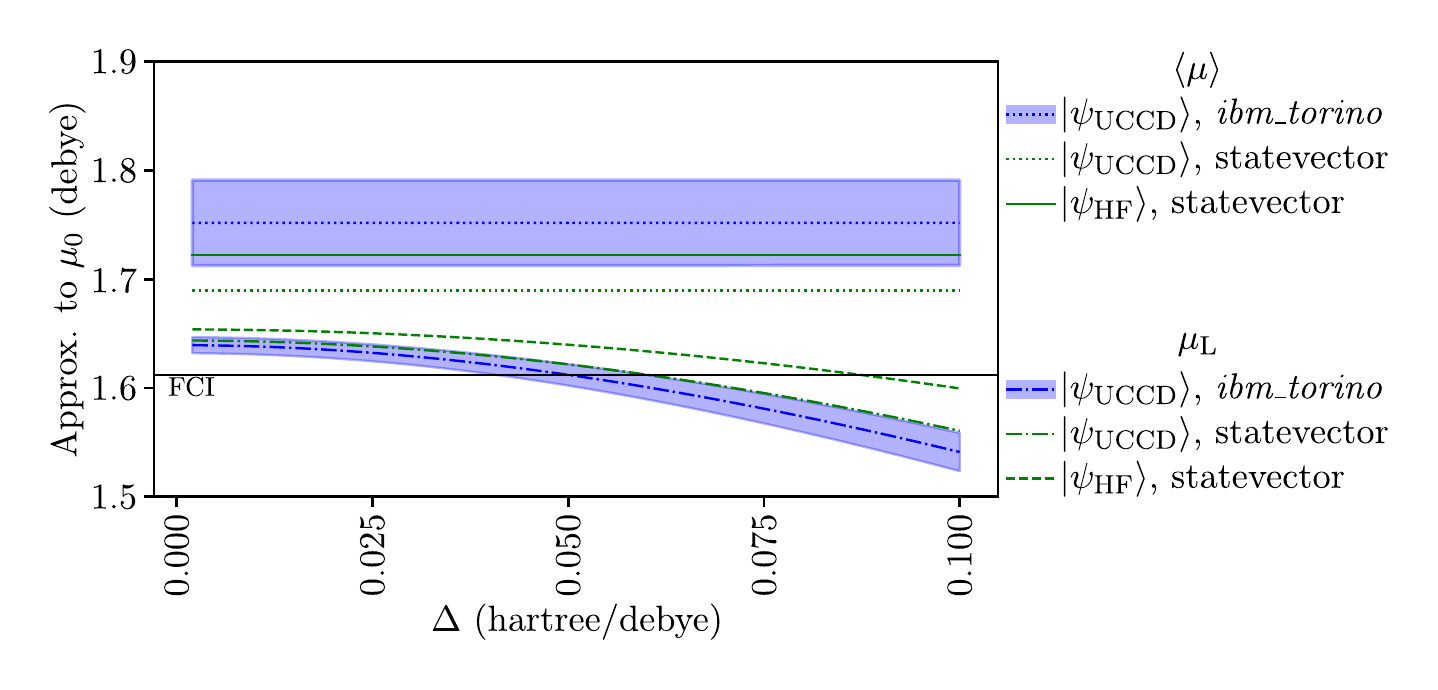}
    \caption{
    Results of evaluating the dipole moment using statevector simulation (green) and quantum hardware (blue) compared with the FCI result (black).
    The dash-dotted and dotted lines indicate the results from the UCCD state with and without the moments-correction respectively. The shaded regions represent 1 standard deviation (determined by 100 bootstrapping samples). Note that as $\Delta$ varies, the approximation is calculated from the same data from the quantum device (or simulator), post-processed using different values of $\Delta$.
    Dashed and solid lines show the results for the Hartree-Fock reference-state with and without the moments correction respectively.
    The finite difference error vanishes in the limit $\Delta\rightarrow0$, the results for larger $\Delta$ values are plotted to demonstrate the convergence of $\mu_\mathrm{L}$.
    }
    \label{fig:Comparison}
\end{figure}

The results of the dipole moment calculation are shown in Figure \ref{fig:Comparison}. With the moments-based correction, the UCCD trial-state results from \emph{ibm\_torino} (blue dash-dotted line) reproduce statevector simulation results (green dash-dotted line) to within 1 standard deviation (shaded area, determined from 100 bootstrapping samples) for small values of $\Delta$, i.e. in the region of interest. On the other hand, direct evaluation of the expectation value (blue dotted line) fails to reproduce the statevector simulation results (green dotted line)\footnote{A guide on reading plotted results: Line patterns indicate whether the moments-correction was used (dashes) and the trial-state (dotted for UCCD). e.g. a dash-dotted line indicates moments-corrected results and the UCCD state, while a solid line (no dashes and no dots) represents no moments and the HF state. Colours correspond to backends; green lines for statevector simulation, and blue for error-mitigated \emph{ibm\_torino} results. Red (appendix only) indicates \emph{ibm\_torino} results without reference-state error mitigation.}.
In addition to the moments-corrected $\mu_\mathrm{L}$ being closer to the target FCI value, its variance is also notably smaller than that of the expectation value, $\langle\mu\rangle$, demonstrating that the observed error robustness property \cite{Vallury_NoiseRobust_2023} of the moments-based energy correction transfers to the evaluation of non-energetic properties.

We note that the expectation values, $\langle\mu\rangle$, appear to be completely independent of $\Delta$, as expected since the finite difference error vanishes when the function is linear as in Equation \ref{Eqn:TrialExpectation}. In truth, there is a very slight $\Delta$ dependence in the error-mitigated (blue dotted) results, since the post-mitigated moments method assigns $\lambda=\Delta$ prior to reference-state calibration, leading to an approximation to the truly linear behaviour of the statevector simulation, however the difference between $\Delta=0$ and $\Delta\rightarrow\infty$ is only 3.7 millidebye. On the other hand $E_\mathrm{L}(\mathcal{H}_\lambda,|\psi\rangle)$ is highly non-linear leading to $\Delta$-dependence in its approximate derivative $\mu_\mathrm{L}$.

While the percentage error in $\mu_\mathrm{L}$ (1.7\%) is smaller than that of $\langle\mu\rangle$ (8.7\%), these are both much larger than the percentage error in the energy estimation (0.05\% and 1.2\% for $E_\mathrm{L}$ and $\langle\mathcal{H}\rangle$ respectively). This is to be expected since for small rotations, $d\theta$, away from the true ground-state the changes in the observables are
\begin{align}
    d\langle\mathcal{H}\rangle_\theta&=\mathcal{O}(d\theta^2),\nonumber\\
    d\langle\mu\rangle_\theta&=\mathcal{O}(d\theta),
\end{align}
where the $\mathcal{O}(d\theta)$ term in the energy change vanishes since the trial-state is near a minimum. The same arguments hold for the changes in the moments $d\langle\mathcal{H}^p\rangle=\mathcal{O}(d\theta^2)$ and modified moments $d\langle\mathcal{H}_\lambda^p\rangle=\mathcal{O}(d\theta)$.

It is also possible that the reason the percentage error in the dipole operator is larger than that of the energy despite having fewer terms in Equation \ref{eqn:Dipole} compared to Equation \ref{eqn:Hamiltonian} is that in the Hamiltonian, the diagonal terms dominate, in particular the 1-body terms, $g_{jj}$, while for the dipole, there is also significant contribution from the off diagonal terms, $f_{jk}$. Since the off-diagonal elements of the reference-state RDM and the mixed-state RDM are the same
\begin{align}
\mathrm{Tr}(\rho^\mathrm{exact}_\mathrm{HF}a^\dagger_ja_k)=\mathrm{Tr}(\rho^\mathrm{mixed}a^\dagger_ja_k)&=0,&j\ne k
\end{align}
they cannot be corrected directly using the reference state error-mitigation since Equation \ref{eqn:NoiseModel} does not have a unique solution for $q$. For this reason, the noise calibration is applied to $\langle\mathcal{H}^p_\Delta\rangle$ instead of the individual RDM elements, which allows error in the off-diagonal elements to be corrected indirectly, essentially by assuming that the correct $q$ value is related to the error level in the diagonal elements. However, a more effective method to correct these off-diagonal elements might be to use trial-states that have non-zero off diagonal terms.

\section{Conclusion}
While ground-state energies are important in the application of quantum algorithms to quantum chemistry, it is also important to be able to evaluate other chemical properties in an accurate, noise-robust way. In contrast to previous results that evaluate expectation values, we have demonstrated that moments-based ground-state energy corrections can be adapted to evaluate such properties (using the electric dipole moment of the water molecule as an example) using the Hellmann-Feynman theorem, and that such a method transfers the inherent accuracy and noise-robustness of the corrected energy value to the quantity of interest.
These observations broaden the potential range of applications in chemistry to which quantum devices could provide a potential advantage over classical computation.

\section{Data Availability}
The data that support the findings of this study are available from the corresponding author upon reasonable request.
\section{Author Contributions}
MAJ and LCLH conceived the project. MAJ set up the computational framework and performed the calculations and data analysis, with input from all authors.
\section{Acknowledgements}
The research was supported by the University of Melbourne through the establishment of the IBM Quantum Network Hub at the University.
This work was financially supported by the Australian Research Council (ARC) Centre of Excellence in Quantum Biotechnology (QUBIC, Grant No. CE230100021).
MAJ and HJV are supported by the Australian Commonwealth Government through Research Training Program Scholarships.
This work was supported by resources provided by the Pawsey Supercomputing Research Centre with funding from the Australian Government and the Government of Western Australia, 
and by The University of Melbourne’s Research Computing Services and the Petascale Campus Initiative. 

\bibliography{bibliography.bib}
\appendix
\setcounter{figure}{0}
\setcounter{equation}{0}
\setcounter{section}{0}
\renewcommand{\thefigure}{A\arabic{figure}}
\renewcommand{\theequation}{A\arabic{equation}}
\renewcommand{\thesection}{A\arabic{section}}
\section{Estimating modified moments in the presence of noise}\label{app:noise}
\subsection{Noise analysis of the pre-mitigated moments method}
\begin{figure}
    \includegraphics[width=\linewidth]{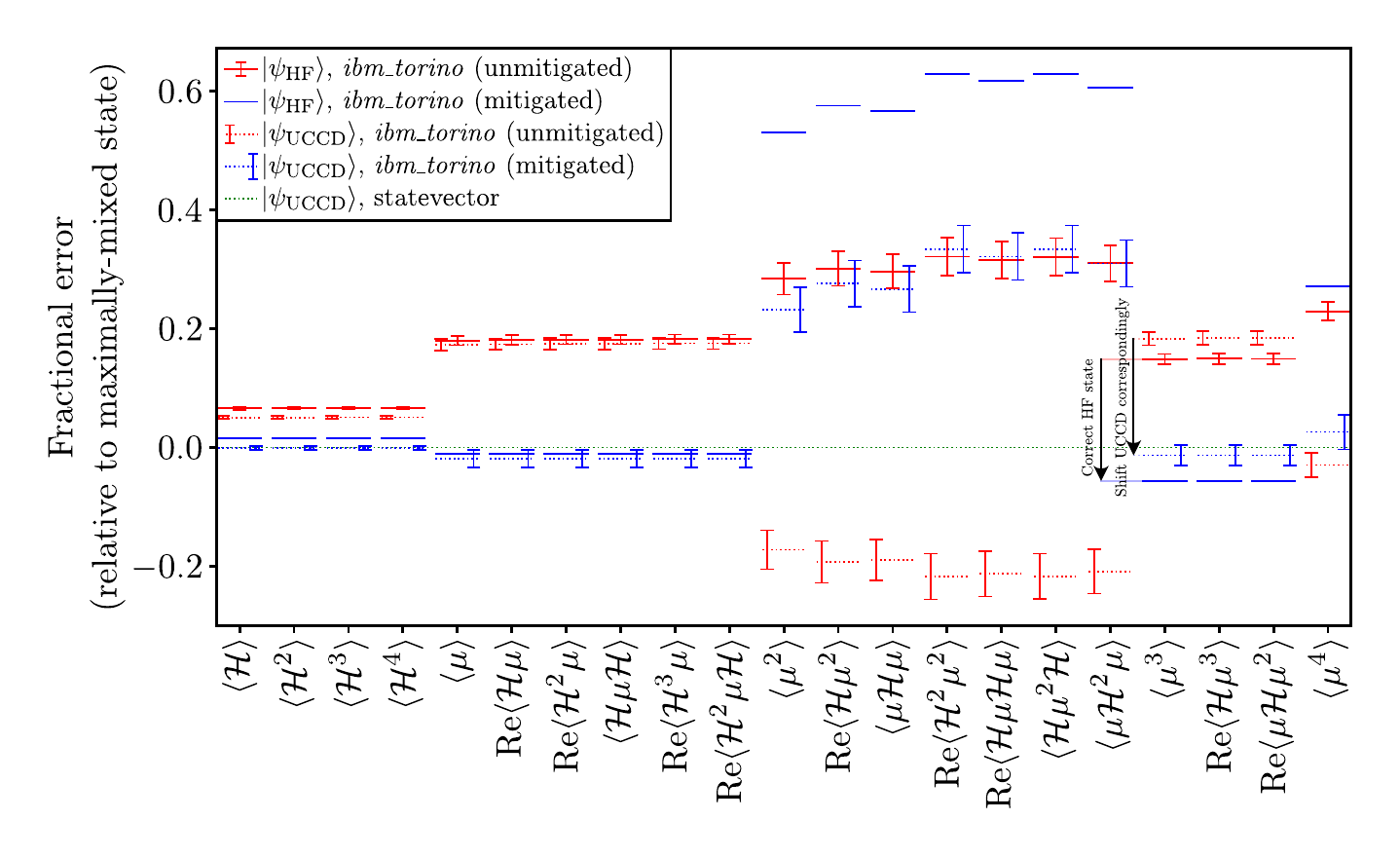}
    \caption{Effectiveness of reference-state error mitigation for the pre-mitigated moments method. The known correction required for the Hartree-Fock state (solid lines) is used to correct the UCCD state results (dotted lines), in both cases the red and blue lines indicate unmitigated and mitigated results respectively. There is no uncertainty in the mitigated Hartree-Fock state values as these are the classically efficient baseline used for the calibration. The errors have been normalised so that the maximally mixed state gives an error of 1. While most of the coefficients have errors on the order of a few percent the terms quadratic in $\mu$ all have error of at least 20\% due to the breakdown of the noise model.
    NB: `unmitigated' means no reference-state error mitigation, other error mitigation techniques have been applied.}
    \label{fig:ReferenceMitigation}
\end{figure}

In Figure \ref{fig:2} we asserted that the pre-mitigated moments method (Method C) was unsuitable due to its unpredictable behaviour in mitigating quantum noise. Here we will investigate this behaviour in more detail. To do this, we will begin at the point where the pre- and post-mitigated methods diverge, that is, with the error mitigation of the coefficients, $\langle\mathcal{H}\rangle$, $\langle\mu\rangle$, $\langle\mathcal{H}\mu\rangle$, etc. Figure \ref{fig:ReferenceMitigation} shows the application of the reference-state error-mitigation technique to these coefficients. Notably the coefficients that are quadratic in the dipole operator, $\mu$, are mitigated poorly. In fact, for quadratic and quartic operators, the noisy measurements drift \emph{away} from the maximally mixed-state compared to their noiseless values. This suggests that the white noise model used in the mitigation does not accurately characterise the behaviour of the device, though it works reasonably well for the unmodified moments. We note that the quartic term is still mitigated reasonably well since the HF and UCCD expectation values are close to each other, relative to the operator norm.

While the fractional errors in Figure \ref{fig:ReferenceMitigation} appear to depend only on the order in $\mu$, there is small variation between terms that is not visible on the scale presented. The final error levels for the UCCD state after mitigation are approximately $-0.1\%\pm0.3\%$ for terms independent of $\mu$, $-2.0\%\pm1.5\%$ for linear terms, $20-30\%\pm4\%$ for the quadratic terms, $-1.3\%\pm1.7\%$ for cubic terms and $2.6\pm2.9\%$ for the quartic term. The error in the final moments-corrected dipole moment estimate (where statevector simulation and mixed state are defined as 0\% and 100\% error respectively) is about $0.1-0.3\%\pm0.2-0.3\%$ depending on the method.

Noting that the error mitigation works well for terms that are constant or linear in $\mu$, and that the dependency on $\Delta$ and $\mu$ matches, the dipole moment estimation should therefore be reasonable in the region $\Delta^2\ll\Delta$. This is observed in Figure \ref{fig:MethodComparison}, which presents the same results as Figure \ref{fig:Comparison} in addition to the equivalent results from Method C and another two Methods (D and E) outlined below.

On the other hand, in Method B, the error mitigation is not applied directly to $\langle\mu^2\rangle$ and is instead applied to $\langle\mathcal{H}_\Delta^2\rangle=\langle\mathcal{H}^2+\Delta\{\mathcal{H},\mu\}+\Delta^2\mu^2\rangle$, where the correction is dominated by the lower order terms in $\Delta$ so effects of the noise model failing are less pronounced. 
If we consider the action of the reference-state error mitigation to be some non-linear function $f$, then the different results between Methods B and C are simply a manifestation of this non-linearity:
\begin{align}
    f\Big(\langle\mathcal{H}\rangle+\Delta\langle\mu\rangle\Big)\ne f\Big(\langle\mathcal{H}\rangle\Big)+\Delta f\Big(\langle\mu\rangle\Big).
\end{align}

\begin{figure}
    \includegraphics[width=\linewidth]{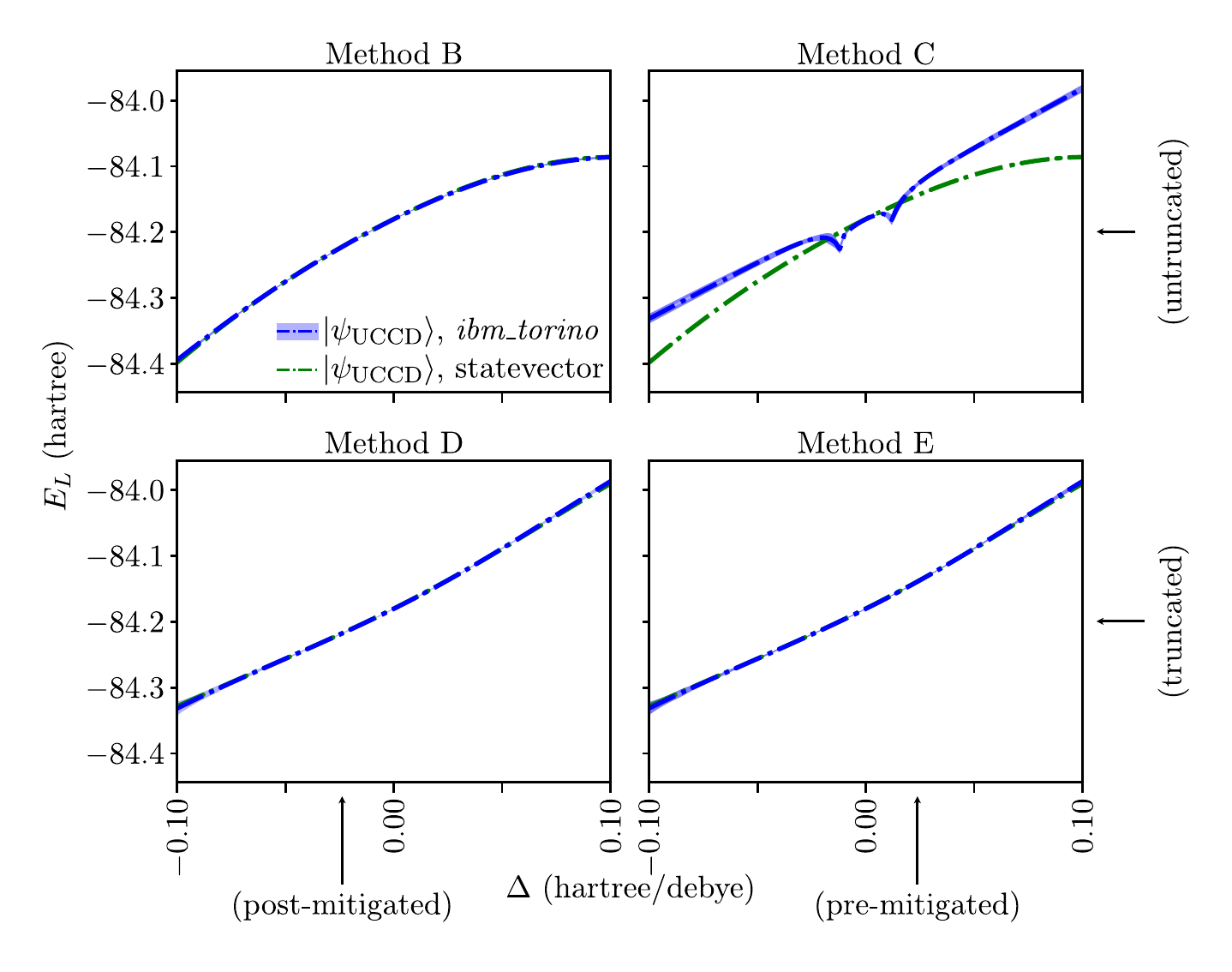}
    \caption{Estimation of $E_\mathrm{L}$ as a function of $\Delta$ through Methods B-E. The error mitigated results from \emph{ibm\_torino} (blue) match the statevector results (green) closely for Methods B, D and E but Method C is too sensitive to noise in the higher order terms in $\Delta$. The left (right) column plots correspond to post- (pre-) mitigated methods and the top (bottom) row to untruncated (truncated) methods respectively.}
    \label{fig:EL}
\end{figure}

It is also curious that in the large $\Delta$ limit, the estimate from pre-mitigation converges to $\langle\mu\rangle$. Examining this, we find that the component of Equation \ref{eqn:EL} inside the square root becomes negative\footnote{This only happens in the presence of significant noise (note the amount by which the mitigation fails in Figure \ref{fig:ReferenceMitigation}) and can serve as an indicator that the noise is too strong.} for $|\Delta|\gtrsim0.01$. Since the imaginary part is discarded the resulting correction term will have the form $-c_3/(c_3^2-c_2c_4)$. In this case, we also see that due to noise, $c_4$ is an order of magnitude larger than $c_2$ or $c_3$ for $|\Delta|\gtrsim0.04$ causing the correction term to vanish and the estimates to converge to the naive expectation value estimate. Additionally, discarding the imaginary component results in a discontinuity in the gradient of $E_\mathrm{L}$. This forms a pair of sharp peaks in Figure \ref{fig:EL} and using points near these in the finite difference approximation leads to the spike observed in Figure \ref{fig:MethodComparison}.

\subsection{Truncation of higher-order terms}
Since the error is predominantly introduced via the quadratic and, to a lesser extent, higher order terms, we could also consider simply discarding these terms. Because the region of interest is the limit as $\Delta\rightarrow0$ this should not affect the final estimate too much.
Running two methods, Methods D and E, corresponding to truncation of Methods B and C respectively, we find that these converge to within 0.1 millidebye of the untruncated versions at $\Delta=0.02$ and expect that the remaining discrepancy will vanish as $\Delta$ approaches 0. Table \ref{tab:method_summary} summarises the differences between Methods A-E.

\begin{table}
    \centering
    \begin{tabular}{|c|c|c|}\hline
        Method&Description&Features\\\hline\hline
        \multirow{3}{*}{A}&&Faster for a single choice\\&$\Delta$ is numeric throughout&of $\Delta$ but need to start over\\&&each time $\Delta$ is changed\\\hline
        \multirow{3}{*}{B}&$\Delta$ is symbolic for moment&Changing $\Delta$ requires the (classical)\\&operator computation and&error calibration to be repeated, but\\&numeric for error calibration&not the moment operator computation\\\hline
        \multirow{3}{*}{C}&&Changing $\Delta$ is trivial, simple\\&$\Delta$ is symbolic throughout&substitution into a formula.\\&&More prone to errors\\\hline
        \multirow{3}{*}{D}&Analogous to B,&Slightly lower cost and\\&but moment operators&very slightly lower\\&are approximated&variance for small $\Delta$\\\hline
        \multirow{3}{*}{E}&Analogous to C,&Slightly lower cost and\\&but moment operators&much lower variance for\\&are approximated&small $\Delta$ (relative to C)\\\hline
    \end{tabular}
    \caption{Brief summary and comparison of moments-corrected dipole moment estimation methods through the Hellmann-Feynman theorem. The key difference between Methods A-C is at what stage the finite difference step-size $\Delta$ is converted from a symbolic parameter to a numeric one. Note that the approximations introduced by Methods D and E converge in the same limit in which the finite difference approximation to the Hellmann-Feynman theorem (present in all Methods) does.}
    \label{tab:method_summary}
\end{table}

\begin{figure}
    \includegraphics[width=\linewidth]{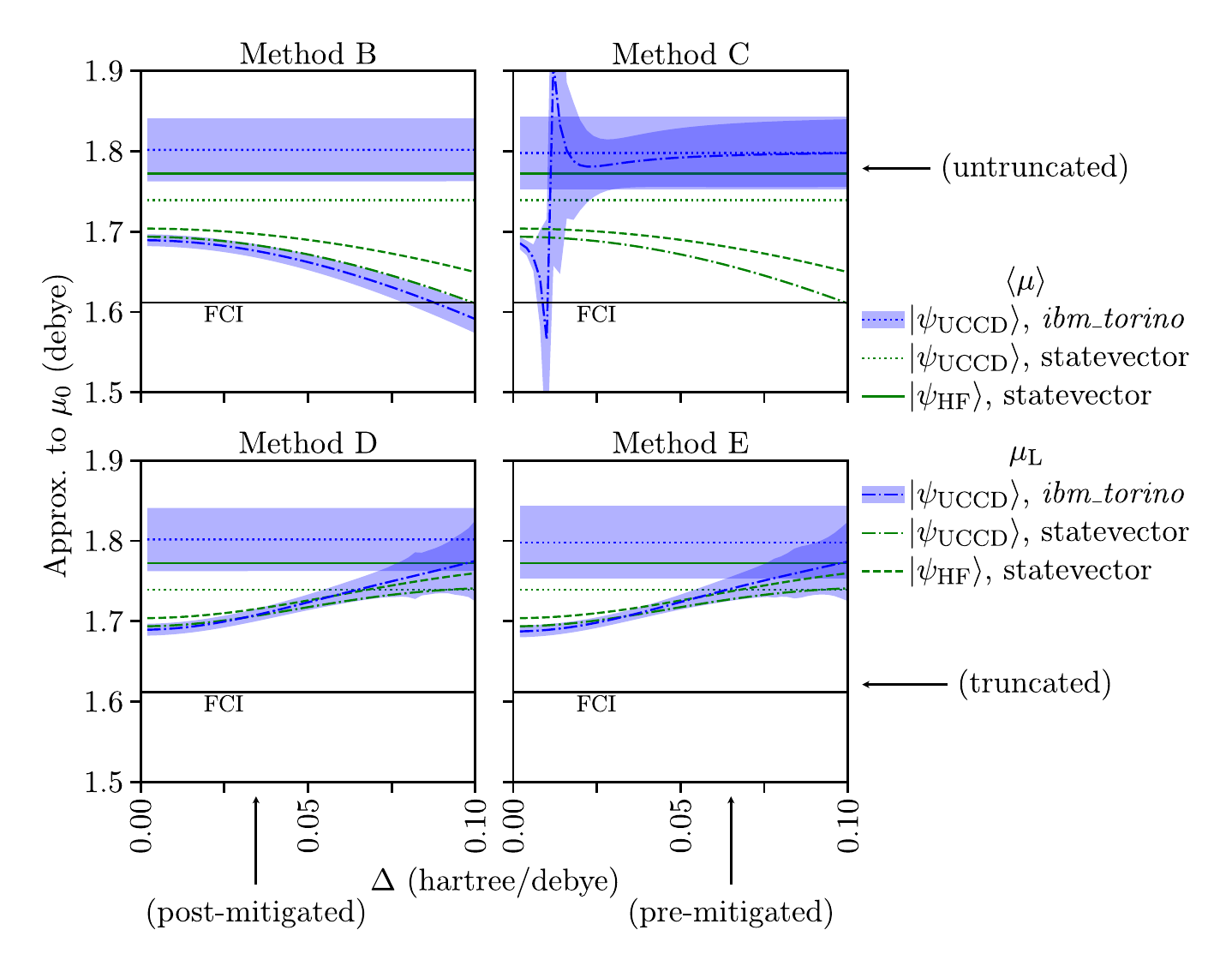}
    \caption{Final estimates of $\mu_0$ using Methods B-E. In each case, the moments-based estimate (blue dash-dotted line) converges (in the $\Delta\rightarrow0$) limit to the statevector result (green dash-dotted line). The blue and green dotted lines indicate the application of the Hellmann-Feynman theorem without the moments-correction for mitigated data and statevector simulation respectively. Results from the Hartree-Fock state are shown in dashed (solid) green with (without) the moments correction. NB. dotted lines indicate UCCD results, dashes represent the moments correction, so dash-dotted lines correspond to both. The colour indicates the backend (hardware or simulator). The left (right) column plots correspond to post- (pre-) mitigated methods and the top (bottom) row to untruncated (truncated) methods respectively.}
    \label{fig:MethodComparison}
\end{figure}

We note in Figure \ref{fig:MethodComparison} that the truncation in Methods D and E affects not only the quantum hardware results, but also the classical simulation values. Due to the truncation of the noisy higher order terms, we find that Method E performs much better than Method C, while Method D offers only a minimal reduction in variance over Method B due to B's much more stable behaviour in comparison to C. We note that if it was the linear terms that had mitigated poorly, then the truncated methods would not be likely to grant any improvement in noise resilience though Method B may still be able to succeed, provided the constant terms are not too noisy.


\end{document}